\documentclass[10pt, conference, letterpaper]{IEEEtran}
\IEEEoverridecommandlockouts

\usepackage{cite}
\usepackage{amsmath,amssymb,amsfonts}
\usepackage{algorithmic}
\usepackage{algorithm}
\usepackage{graphicx}
\usepackage{textcomp}
\usepackage{xcolor}
\usepackage{booktabs}
\usepackage{multirow}
\def\BibTeX{{\rm B\kern-.05em{\sc i\kern-.025em b}\kern-.08em
    T\kern-.1667em\lower.7ex\hbox{E}\kern-.125emX}}
\begin{document}

\title{AceSpec: An Asymmetric Edge-Cloud Collaborative Framework for Communication-Efficient LLM Inference}

\author{\IEEEauthorblockN{Yida Zhang}
\IEEEauthorblockA{\textit{School of Computer and Communication Engineering} \\
\textit{University of Science and Technology Beijing}\\
Beijing, China \\
ustbzyd@163.com}
\and
\IEEEauthorblockN{Zhiyong Gao}
\IEEEauthorblockA{\textit{School of Computer and Communication Engineering} \\
\textit{University of Science and Technology Beijing}\\
Beijing, China \\
gzy@xs.ustb.edu.cn}
\and
\IEEEauthorblockN{Shuaibing Yue}
\IEEEauthorblockA{\textit{School of Computer and Communication Engineering} \\
\textit{University of Science and Technology Beijing}\\
Beijing, China \\
m202421086@xs.ustb.edu.cn}
\and
\IEEEauthorblockN{Jie Li}
\IEEEauthorblockA{\textit{School of Computer and Communication Engineering} \\
\textit{University of Science and Technology Beijing}\\
Beijing, China \\
m202521017@xs.ustb.edu.cn}
\and
\IEEEauthorblockN{Rui Wang}
\IEEEauthorblockA{\textit{School of Computer and Communication Engineering} \\
\textit{University of Science and Technology Beijing}\\
Beijing, China \\
wangrui@ustb.edu.cn}
}

\maketitle

\begin{abstract}
Deploying Large Language Models (LLMs) on edge devices typically relies on model compression or split inference. However, compression degrades reasoning capabilities, while split inference suffers from severe Wide Area Network (WAN) communication bottlenecks. Edge-cloud speculative decoding emerges as a promising alternative, leveraging an edge small model to draft tokens for cloud verification. Yet, over volatile WANs, inevitable prediction rejections trigger catastrophic pipeline stalls and network-wide rollbacks, neutralizing collaborative gains. To overcome this, we propose AceSpec, an asymmetric edge-cloud collaborative framework. AceSpec utilizes un-saturated edge compute to proactively construct a probabilistic state cache, effectively transforming network-wide pipeline flushes into $\mathcal{O}(1)$ local memory lookups. To preserve bandwidth, it employs an asymmetric communication protocol that transmits minimal main-chain indices uplink and compact sparse distributions downlink. Furthermore, we introduce a network-aware, Lagrangian-optimized resource allocation strategy that dynamically maximizes the local cache hit rate. Evaluations demonstrate that AceSpec achieves up to a 3.52$\times$ throughput speedup and exhibits exceptional bandwidth immunity, sustaining near-peak inference performance even under severely constrained 50 Kbps WAN conditions.
\end{abstract}


\section{Introduction}
\label{sec:introduction}

Deploying Large Language Models (LLMs) directly at the network edge is increasingly necessary to support applications requiring advanced reasoning. This edge-native approach minimizes end-to-end interaction latency, enables real-time responsiveness. However, the high memory and computational requirements make pure on-device execution impractical for resource-constrained edge nodes. To address this, traditional methods primarily rely on model compression or edge-cloud split inference. Unfortunately, aggressive compression techniques inevitably degrade the reasoning and generative capabilities of the model\cite{zheng2025review}. Split inference partitions the neural network across the edge and cloud, but transmitting large intermediate hidden states over Wide Area Networks (WANs) introduces a ``Communication Wall''. Under low-bandwidth and high-latency WAN conditions, this heavy transmission payload outweighs the computational acceleration, making split inference inefficient\cite{wang2025empowering}.

To avoid the heavy tensor transmission of split inference, speculative decoding \cite{chen2023accelerating} has emerged as a promising alternative, compressing the communication payload from continuous dense tensors to discrete token sequences. By using a lightweight Small Language Model (SLM) at the edge to draft candidate tokens for parallel verification by a cloud-hosted target LLM, this paradigm significantly reduces data volume. However, vanilla speculative decoding was fundamentally designed for intra-node environments or server clusters equipped with high-speed interconnects (e.g., NVLink or PCIe), implicitly assuming negligible communication overhead. Consequently, when directly deployed over high-latency Wide Area Networks (WANs) in edge-cloud collaborative settings, its synchronous ``stop-and-wait'' dependency leads to severe communication bottlenecks\cite{wang2025comprehensive}. To bridge this gap and hide the long WAN Round-Trip Time (RTT), recent frameworks attempt asynchronous pipelines. For instance, PicoSpec \cite{zhang2026pipelined} decouples the edge-cloud dependency using a single-chain asynchronous pipeline, PipeSD \cite{han2026pipesd} introduces token-batch scheduling to stream drafts, and $E^2$-SCI \cite{Li_2026_CVPR} employs progressive lookahead concurrency with adaptive verification thresholds.

Despite mitigating partial latency, these architectures exhibit significant limitations when deployed over volatile, low-bandwidth WANs. Pipelined frameworks like PicoSpec rely on a tightly coupled, single-chain state synchronization mechanism. While PicoSpec preserves generative diversity by supporting stochastic nucleus sampling, the inherent distribution divergence between the draft and target models leads to frequent prediction rejections. Under a single-chain paradigm, these rejections cause pipeline stalls, resulting in substantial network-wide rollback penalties. In contrast, subsequent frameworks like PipeSD and $E^2$-SCI dropping support for nucleus sampling. Supporting nucleus sampling traditionally necessitates transmitting vocabulary-wide draft probability distributions over the uplink, which readily saturates constrained WAN bandwidth. As a result, these newer frameworks fall back to deterministic greedy decoding. Even within this restricted scope, they introduce new inefficiencies. The batch transmission of PipeSD is highly vulnerable to early token rejections, where a single misprediction renders the unverified remainder of the batch obsolete and exacerbates bandwidth underutilization. Meanwhile, $E^2$-SCI relies on frequent control signaling, which inflates protocol overhead without effectively overcoming the physical RTT lower bound. Furthermore, both frameworks require additional fine-tuning or the integration of auxiliary networks, thereby reducing their universality and deployment flexibility for off-the-shelf LLMs at the edge.

To overcome this communication barrier and provide a general, training-free inference solution over WANs, we propose AceSpec, an Asymmetric Collaborative Edge-Cloud Speculative Decoding framework. By trading the abundant parallel compute of the edge device for high WAN rollback costs, AceSpec proactively expands a probabilistic multi-branch token tree cache during the asynchronous verification window. To maximize this latency-masking efficacy, we manage the cache expansion through a two-stage analytical optimization. Network-Aware Budgeting bounds the total compute budget to hide local tree construction within the WAN RTT, and Task-Aware Shape Allocation uses a Lagrangian-optimized, non-uniform geometric decay strategy to deploy branches precisely where prediction rejections are most probable. Coupled with an asymmetric transmission protocol that uploads minimal main-chain indices and downloads sparsified target distributions upon rejection, AceSpec dynamically queries its optimized cache via local resampling. This transforms costly network-wide pipeline flushes into $\mathcal{O}(1)$ local memory lookups, eliminating rollback penalties while remaining universally applicable to any off-the-shelf LLM.

The main contributions of this paper are summarized as follows:
\begin{itemize}
    \item We propose AceSpec, a training-free, asymmetric edge-cloud collaborative architecture that natively supports stochastic nucleus sampling. AceSpec transforms catastrophic WAN rollbacks into $\mathcal{O}(1)$ local memory lookups without inflating uplink traffic.
    \item We formulate a two-level constrained optimization framework to systematically balance edge compute and network delays. It establishes a bandwidth-aware computational bound coupled with a Lagrangian-optimized, non-uniform geometric allocation strategy to strictly maximize the local state cache hit rate.
    \item Extensive evaluations on NVIDIA AGX Orin and A100 clusters demonstrate that AceSpec provides robust bandwidth immunity. It achieves up to a 3.52$\times$ throughput speedup over traditional baselines, sustaining highly efficient generation even under severely constrained WAN conditions (down to 50 Kbps).
\end{itemize}

\section{Related Work}

\subsection{Speculative Decoding}

Speculative decoding accelerates LLM inference by substituting serial generation with a ``draft-then-verify'' paradigm. Chen et al. \cite{chen2023accelerating} demonstrated that a modified rejection sampling mechanism can guarantee the output remains mathematically identical to the target model while providing significant speedups. SpecInfer \cite{miao2024specinfer} introduced a token tree verification mechanism, enabling the target model to evaluate multiple candidate branches concurrently. To enhance draft quality and throughput, Medusa \cite{10.5555/3692070.3692273} employs multiple parallel decoding heads fine-tuned on a frozen backbone, while EAGLE \cite{10.5555/3692070.3693232} uses feature-layer extrapolation to generate high-quality drafts with minimal overhead. Recent studies further optimize drafting costs and parallelization; for instance, PEARL \cite{liu2025pearl} and SwiftSpec \cite{zhang2026swiftspec} explore advanced verification mechanisms and parallel tree generation, while CoSine \cite{gao2025collaborative} routes requests across multiple GPUs. Furthermore, SSD \cite{kumar2026speculative} pre-emptively prepares a speculation cache for multiple verification outcomes to eliminate local drafting latency.

However, these methods implicitly assume near-zero communication costs and are fundamentally designed for intra-node environments or datacenter servers equipped with high-speed interconnects (e.g., NVLink or PCIe). When directly applied to edge-cloud collaborative environments, they exhibit significant performance degradation. The core issue lies in the mismatch between the high-latency, bandwidth-constrained nature of Wide Area Networks (WANs) and the frequent state-synchronization required by the iterative draft-then-verify paradigm. Consequently, they cannot be deployed directly in WAN environments without exacerbating the communication bottleneck.

\subsection{Edge-Cloud Inference}
Edge-cloud collaborative inference aims to overcome the resource constraints of edge nodes through model splitting and task offloading. Frameworks like EdgeShard \cite{zhang2024edgeshard} and Jupiter \cite{ye2025jupiter} employ dynamic programming and intra-sequence pipeline parallelism to optimize model splitting across distributed nodes. Other research focuses on hybrid routing: MMSL \cite{ma2025multi} uses multi-stage scheduling, while Hybrid LLM \cite{ding2024hybrid} and Hybrid SLM and LLM \cite{hao2024hybrid} route specific requests or low-confidence tokens to the cloud based on task difficulty. 

Despite reducing the local computational load, these spatial or routing-based schemes fail to overcome the ``Communication Wall.'' The edge device must idle during remote processing and data transmission, tying the inference speed tightly to the physical limits of network RTT and restricted uplink bandwidth. Consequently, the theoretical acceleration of offloading is often offset by the communication overhead in WANs.

\subsection{Distributed Speculative Decoding}

Recent efforts have attempted to adapt speculative decoding for distributed environments. To optimize edge performance, EdgeLLM \cite{xu2024edgellm} employs local parallel tree generation but imposes heavy memory and computing burdens on limited hardware without considering bandwidth efficiency. Other frameworks like Venkatesha et al. \cite{venkatesha2025fast} and HAT \cite{xie2025novel} introduce early exits or hidden-state exchanges; however, they often require invasive model modifications or inflate transmission payloads, easily saturating WAN bandwidth. To reduce uplink data, DSSD \cite{ning2025dssd} moves re-sampling back to the local edge. Frameworks like DSD \cite{yu2025dsd} and SLED \cite{li2025sled} also explore dynamic windows and shared serving. Fundamentally, these architectures still adhere to a synchronous ``stop-and-wait'' drafting paradigm, leaving the edge idle during long WAN round trips. 

To further reduce communication delays, recent frameworks propose pipelined architectures. PicoSpec \cite{zhang2026pipelined} decouples the edge-cloud dependency using a single-chain asynchronous pipeline. PipeSD \cite{han2026pipesd} overlaps token generation and communication via a token-batch scheduling mechanism, while $E^2$-SCI \cite{Li_2026_CVPR} introduces progressive lookahead concurrency with adaptive verification thresholds. Despite mitigating partial network delays, these pipelines exhibit critical systemic bottlenecks in practical low-bandwidth deployments. To prevent large vocabulary-wide probability tensors from saturating the WAN uplink, they frequently fall back to deterministic greedy sampling, which prevents them from supporting the stochasticity required by real-world generative tasks. Furthermore, their reliance on single-chain drafting makes them highly vulnerable. When prediction rejections inevitably occur under nucleus sampling, the single sequence is invalidated, forcing the system to absorb substantial network-wide rollback delays. While PipeSD and $E^2$-SCI attempt to bypass this, they either exacerbate bandwidth waste via obsolete batch transmissions or inflate protocol overhead, and they often require invasive model retraining. Although naive tree-based speculation could theoretically improve acceptance rates, it is generally considered unsuitable for WAN settings due to its prohibitive communication overhead \cite{han2026pipesd}.

These limitations underscore the need for AceSpec: a training-free, asymmetric architecture that uses available local compute to proactively build probabilistic multi-branch caches. By utilizing optimal resource allocation to hide WAN latency, AceSpec effectively eliminates rollback penalties without inflating communication overhead.

\section{Method}

\subsection{The Rollback Bottleneck in Edge-Cloud Inference Pipelines}
\label{subsec:bottleneck}

To achieve efficient LLM inference on resource-constrained edge devices, edge-cloud collaboration via speculative execution is widely adopted. In this setup, an edge device uses a local Small Language Model (SLM) to generate a draft of length $\gamma$, while a cloud server performs parallel verification using a Target LLM. Let $\alpha \in (0, 1)$ denote the average single-token acceptance rate. The expected number of accepted tokens per round is $\mathbb{E}[L] = (1-\alpha^\gamma)/(1-\alpha)$.

In vanilla deployments, this architecture operates in a synchronous ``stop-and-wait'' paradigm. The cycle time is $T_{naive} = T_{draft}(\gamma) + RTT + T_{verify}$, where $RTT$ is the Network Round-Trip Time and $T_{verify}$ is the cloud verification time. In Wide Area Networks (WANs) where the $RTT$ is high, this synchronous dependency creates a ``Communication Wall'' that leaves edge compute underutilized and significantly limits inference throughput.

To mitigate this bottleneck and maximize edge computing efficiency, existing frameworks attempt to overlap edge computation with network transmission using asynchronous pipelines. Ideally, if the cloud accepts every draft, the WAN delay is fully masked, yielding an optimal cycle time $T_{ideal} = \max(T_{draft}(\gamma), RTT + T_{verify})$. However, this approach relies on a vulnerable single-chain state synchronization mechanism. The probability of a full sequence acceptance decays exponentially ($p_{hit}^{single} = \alpha^\gamma$). 

When a prediction rejection occurs, the asynchronous pipeline stalls, introducing a substantial pipeline bubble ($T_{bubble}$) consisting of downlink transmission, sequential edge redrafting, and uplink transmission delays. We formulate the expected time penalty incurred by these frequent network-wide pipeline flushes as:
\begin{equation}
    \mathbb{E}[Penalty] = (1 - \alpha^\gamma) \times (RTT + T_{redraft}(\gamma))
    \label{eq:penalty}
\end{equation}

By incorporating this penalty, the realistic throughput of pipelined edge-cloud inference reduces to:
\begin{equation}
    R_{pipe} = \frac{\mathbb{E}[L]}{T_{ideal} + \mathbb{E}[Penalty]}
    \label{eq:throughput}
\end{equation}

Equation (\ref{eq:throughput}) highlights a fundamental limitation of current edge-cloud inference pipelines. As the acceptance rate $\alpha$ decreases, the penalty coefficient $(1 - \alpha^\gamma)$ approaches 1. The pipeline bubble ($RTT + T_{redraft}$) dominates the denominator, forcing the system's cycle time to revert toward the synchronous baseline $T_{naive}$. Ultimately, without addressing this network rollback penalty, edge devices cannot achieve high-efficiency inference because their local compute capabilities are frequently stalled by WAN volatility.

\subsection{AceSpec}
\label{subsec:lifecycle}

\begin{figure}[t]
    \centering
    \includegraphics[width=\linewidth]{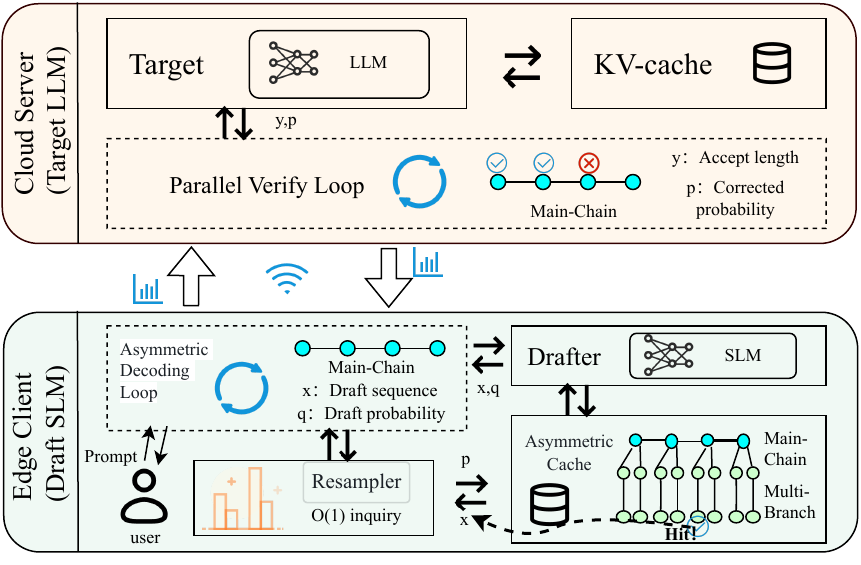}
    \caption{The AceSpec System Architecture.}
    \label{fig:architecture}
\end{figure}

To overcome communication bottlenecks and prevent pipeline stalls, AceSpec introduces an asymmetric edge-cloud collaborative architecture. The core principle of this design is to shift the computational and decision-making burdens toward the available parallel compute of the edge, thereby minimizing the volume of data exchanged over the WAN. As illustrated in the system architecture (Fig. \ref{fig:architecture}), the system comprises two primary operational entities: an Edge Client and a Cloud Server. The Edge Client hosts the lightweight draft SLM and maintains three integrated subsystems: the Tree-based Drafter, the Probabilistic KV Cache Manager, and the Local Resampler. These components interact to generate alternative speculative paths, dynamically structure the multi-branch token tree in memory, and execute non-blocking network I/O. The Cloud Server hosts the high-parameter target LLM and executes a highly parallelized Verify Loop. Rather than controlling the final token selection, the cloud acts as an evaluator that checks the validity of the edge's proposals and generates probability distributions upon verification failures.

\begin{figure*}[t]
    \centering
    \includegraphics[width=\linewidth]{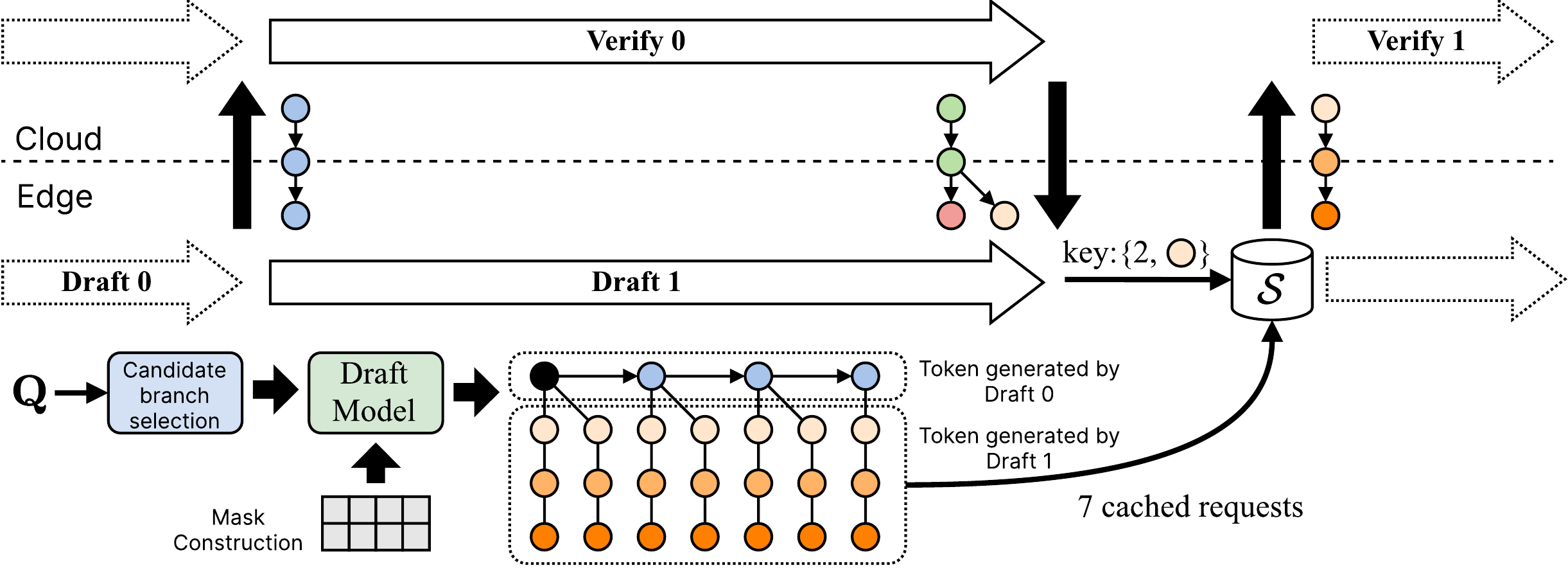}
    \caption{The Execution Pipeline of AceSpec.}
    \label{fig:pipeline}
\end{figure*}

The execution lifecycle of AceSpec operates as a continuous, overlapping loop characterized by asynchronous data flows and localized state synchronization. To demonstrate how these architectural components interact, Fig. \ref{fig:pipeline} maps out the temporal execution timeline. After completing a drafting iteration (e.g., Draft 0), the edge client executes an asymmetric stream transmission. Instead of transmitting the entire generated multi-branch tree to the cloud, it extracts only the primary linear sequence (the main chain) and sends the token indices to the server. This protocol reduces the uplink communication volume to a scalar array, conserving the constrained WAN uplink bandwidth.

After dispatching the main chain, the edge client uses the cloud turnaround time $T_{cloud}$ as a computation window. The Tree-based Drafter performs an asynchronous multi-branch expansion starting from potential rejection points across the main chain. For instance, Draft 1 expands concurrently with the cloud's Verify 0 stage, allowing the edge to store the resulting state sequences in the Probabilistic Cache $\mathcal{S}$.

Concurrently, the cloud server processes the received main chain tokens in a single forward pass. If the target LLM rejects a proposed token at a specific position, the cloud server halts further validation and skips token resampling. Instead, it packages the rejection index with the corresponding partial target probability distribution for that position and returns this feedback to the edge.

Upon receiving this feedback, the Local Resampler on the edge client processes the partial target distribution and performs local token resampling to determine the correct token. This resolved token is then used as a query key (e.g., \texttt{key: \{2, O\}}) to search the local tree cache. If the token matches a pre-computed alternative branch, the Probabilistic KV Cache Manager immediately updates the active context pointer. This local state recovery bypasses the sequential redrafting penalty, effectively hiding the WAN latency and initiating the next pipeline cycle.

\subsection{Asymmetric Tree Construction and $\mathcal{O}(1)$ State Recovery}
\label{subsec:recovery}

\begin{figure}[t]
    \centering
    \includegraphics[width=\linewidth]{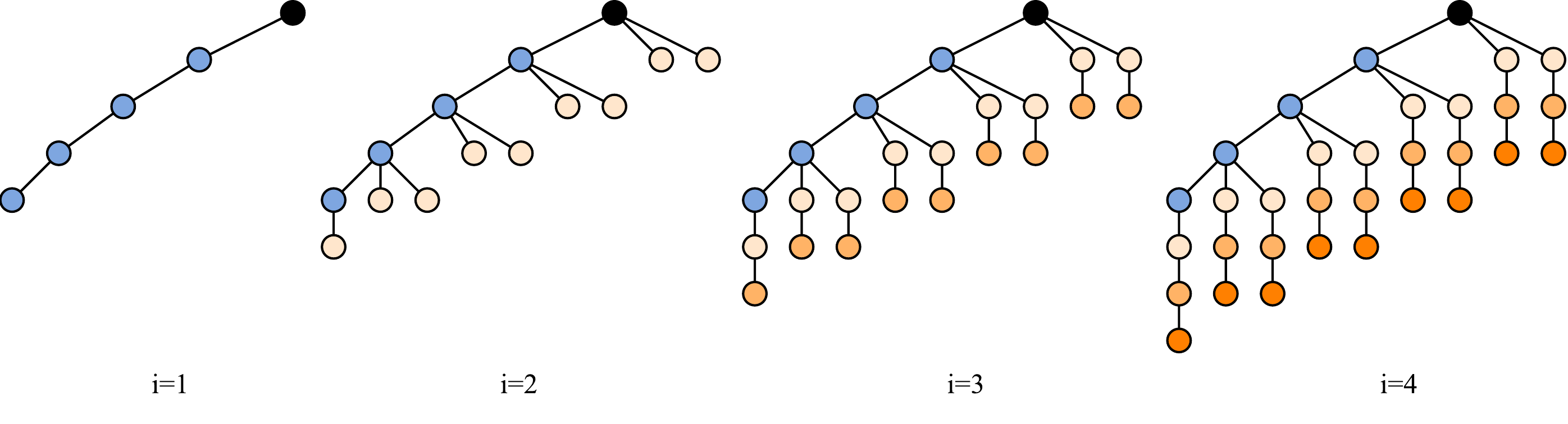}
    \caption{The progressive growth of the multi-branch token tree across different decoding steps ($i=1$ to $4$) during the asynchronous window.}
    \label{fig:tree_growing}
\end{figure}

To implement the asynchronous pipeline introduced in Section \ref{subsec:lifecycle}, AceSpec decouples local edge computation from network constraints using Asymmetric Branch Caching. During the cloud turnaround window ($T_{cloud}$), the edge SLM uses its idle compute to expand a multi-branch token tree $\mathcal{T}$. As illustrated in Fig. \ref{fig:tree_growing}, this tree grows progressively across different decoding steps. Because the target distribution $P_{k+1}$ of the cloud is unknown a priori, the edge uses its local draft distribution $Q_{k+1}$ as a proxy. At each potential rejection depth $k \in \{0, \dots, \gamma-1\}$, it excludes the main-chain token and extracts the Top-$F_k$ candidate tokens from the residual probability mass to construct alternative branches.

\begin{figure}[t]
    \centering
    \includegraphics[width=\linewidth]{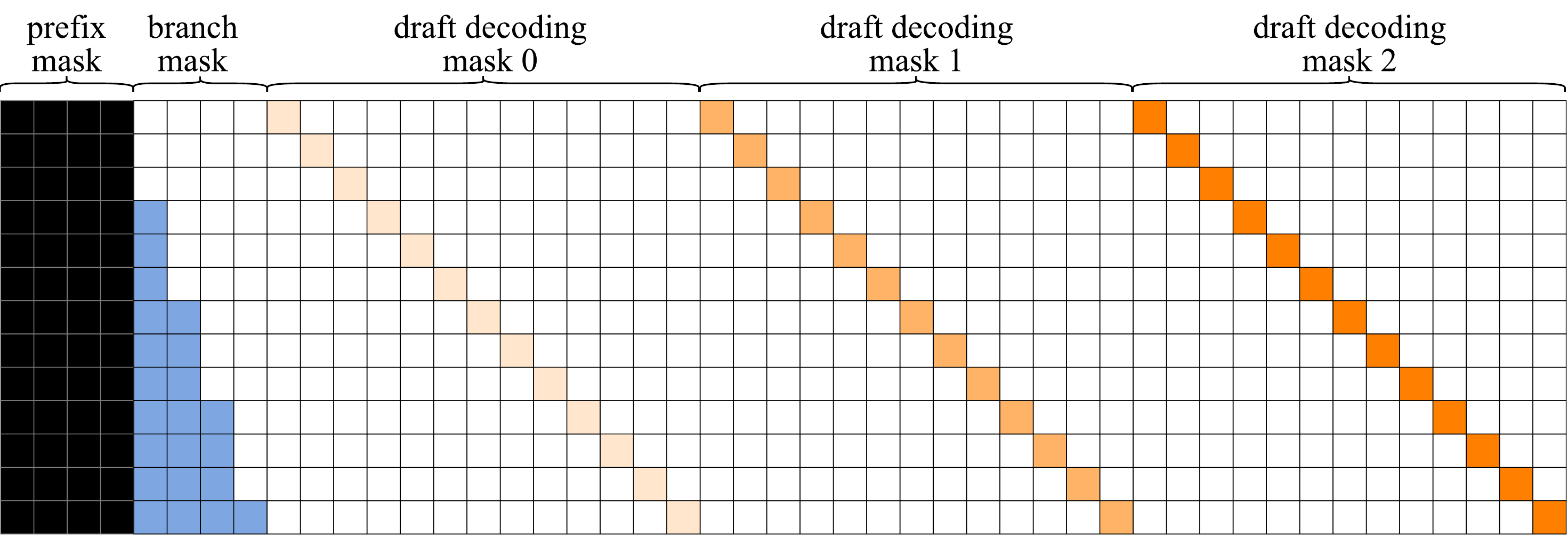}
    \caption{The topology of the Tree Attention Mask, comprising the Shared Prefix Block, Main-Path Branch Block, and Suffix Diagonal Block.}
    \label{fig:tree_mask}
\end{figure}

To prevent drafting latency from scaling linearly with the total number of branches $N$, AceSpec implements a customized Tree Attention Mask. Instead of executing $N$ sequential forward passes, all branches are packed into a single parallel forward pass. As detailed in Fig. \ref{fig:tree_mask}, the mask comprises three distinct sub-blocks. The Shared Prefix Block allows all $N$ branches to attend to the historically verified context $x_{1:t}$. The Main-Path Branch Block constrains a branch originating from depth $s_j$ to attend only to the first $s_j$ tokens of the main draft. Finally, the Suffix Diagonal Block enforces a diagonal isolation constraint so each speculative token attends exclusively to its own ancestral sequence. Consequently, the multi-branch expansion completes within an $\mathcal{O}(1)$ wall-clock duration.

\begin{algorithm}[t]
\caption{AceSpec Edge-Side Asynchronous Pipeline}
\label{alg:acespec_edge}
\begin{algorithmic}[1]
\REQUIRE Draft SLM $\mathcal{M}_{edge}$, Draft length $\gamma$, Fan-out vector $F$, Input prefix $X$.
\STATE Initialize KV Cache and generate initial draft $\tilde{x}, Q \leftarrow \text{DRAFT}(\mathcal{M}_{edge}, X, \gamma)$
\WHILE{generation not finished}
    \STATE $\hat{q} \leftarrow \text{COMPRESS}(Q)$ \COMMENT{Asymmetric Transmission}
    \STATE $\text{SEND\_ASYNC}(\text{Cloud}, \langle \tilde{x}, \hat{q}, \gamma \rangle)$
    \STATE $\mathcal{T} \leftarrow \text{TREE\_DRAFT}(\mathcal{M}_{edge}, \tilde{x}, Q, F)$ \COMMENT{Asynchronous Multi-Branching}
    \STATE $(\pi, \hat{P}_{target}) \leftarrow \text{RECEIVE\_WAIT}(\text{Cloud})$ \COMMENT{Wait for verification}
    \STATE $x_{resampled} \leftarrow \text{RESAMPLE}(\hat{P}_{target}, Q)$ \COMMENT{Local Token Resampling}
    \STATE $w \leftarrow (\pi, x_{resampled})$ \COMMENT{Verification Outcome}
    \IF{$w \in \text{dom}(\mathcal{T})$}
        \STATE $(\tilde{x}_{next}, Q_{next}) \leftarrow \mathcal{T}[w]$ \COMMENT{$\mathcal{O}(1)$ State Recovery}
        \STATE $\text{COMPACT\_KV}(\mathcal{M}_{edge}, w)$ \COMMENT{Continuous memory alignment}
        \STATE $\tilde{x} \leftarrow \tilde{x}_{next}; Q \leftarrow Q_{next}$
    \ELSE
        \STATE $\text{ROLLBACK\_KV}(\mathcal{M}_{edge}, \gamma - \pi)$ \COMMENT{Cache miss penalty}
        \STATE $\tilde{x}, Q \leftarrow \text{DRAFT}(\mathcal{M}_{edge}, x_{resampled}, \gamma)$ \COMMENT{Sequential Redrafting}
    \ENDIF
\ENDWHILE
\end{algorithmic}
\end{algorithm}

The execution logic of this state recovery is formalized in Algorithm \ref{alg:acespec_edge}. After receiving the sparsified target distribution $\hat{P}_{target}$ and acceptance length $\pi$, the edge computes the corrected token $x_{resampled}$ via a local resampling operation (Line 7). The state transition occurs at Line 9. Rather than idling for a costly sequential redraft ($T_{redraft}$), the edge uses the outcome $w = (\pi, x_{resampled})$ as a query key to look up the locally cached tree $\mathcal{T}$. Upon a cache hit, it retrieves the pre-computed draft and executes a \texttt{COMPACT\_KV} operation (Line 11). By using memory scatter/gather primitives, this operation aligns the KV entries of the hit branch behind the verified prefix and invalidates unselected branches. This local $\mathcal{O}(1)$ memory compaction eliminates the sequential redrafting penalty, transforming a costly pipeline flush into a zero-overhead local memory lookup.

\section{Theoretical Optimal Branch Allocation}
\label{sec:allocation}

\subsection{Expected Time and Speedup Analysis of Multi-Branch Caching}

To analyze the performance gain, we define the theoretical speedup $S$ of AceSpec relative to the standard synchronous baseline. We abstract the latency components into three system constants. Let $T_{naive} = T_{draft}(\gamma) + RTT + T_{verify}$ denote the cycle time of the stop-and-wait baseline. For the pipelined architecture, we define $T_{ideal} = \max(T_{tree\_draft}(B), RTT + T_{verify})$ as the ideal cycle time assuming a perfect cache hit rate, and $T_{bubble} = RTT + T_{redraft}(\gamma)$ as the time penalty incurred upon a cache miss.

Using these variables, the expected cycle time of the multi-branch system is expressed as $\mathbb{E}[T_{tree}] = T_{ideal} + (1 - P_{hit}^{tree}(\mathcal{T})) T_{bubble}$. Because the multi-branch tree provides at least the same expected accepted length $\mathbb{E}[L]$ as a single chain, the lower bound of the speedup $S$ is given by:

\begin{equation}
    S = \frac{R_{tree}}{R_{naive}} \ge \frac{T_{naive}}{T_{ideal} + (1 - P_{hit}^{tree}(\mathcal{T})) T_{bubble}}
    \label{eq:speedup_lower_bound}
\end{equation}

By dividing the numerator and denominator by $T_{ideal}$, we derive a simplified formulation:
\begin{equation}
    S \ge \frac{S_{ideal}}{1 + (1 - P_{hit}^{tree}(\mathcal{T})) \cdot \rho}
    \label{eq:speedup_simplified}
\end{equation}
where $S_{ideal} = T_{naive}/T_{ideal}$ represents the maximum theoretical speedup limit when the pipeline is perfectly masked, and $\rho = T_{bubble}/T_{ideal}$ represents the Rollback Penalty Ratio dictated by the WAN environment and hardware compute constraints.

Equation \eqref{eq:speedup_simplified} separates the physical limits of the system from its algorithmic performance.

\subsection{Problem Formulation with Two-Level Constraints}

As established in Eq. \eqref{eq:speedup_simplified}, maximizing the tree cache hit rate $P_{hit}^{tree}(\mathcal{T})$ is a prerequisite for hiding latency. However, constructing $\mathcal{T}$ is bounded by two distinct constraints: network conditions and algorithmic properties. To maximize $P_{hit}^{tree}(\mathcal{T})$, we formulate a two-level constrained optimization problem.

\subsubsection{Network-Aware Budgeting}
The edge device cannot arbitrarily increase the tree size $B$ to improve the hit rate. To ensure multi-branch drafting does not violate the ideal pipeline boundary and become a new computational bottleneck, the edge execution time must be masked by the cloud turnaround time: $T_{tree\_draft}(B) \le RTT(t) + T_{verify}$. 

By continuously probing the real-time WAN delay $RTT(t)$, AceSpec derives a network-aware computation budget $B_{net}(t) = \lfloor T_{tree\_draft}^{-1}(RTT(t) + T_{verify}) \rfloor$, where $T_{tree\_draft}^{-1}(\cdot)$ is the inverse latency function of the edge SLM. Considering the physical VRAM and parallel capacity limit of the edge device ($B_{max}$), the effective budget $B^*(t)$ is dynamically bounded as:

\begin{equation}
    B^*(t) = \min \big( B_{max}, \lfloor T_{tree\_draft}^{-1}(RTT(t) + T_{verify}) \rfloor \big)
    \label{eq:budget}
\end{equation}

This bounding provides a compute-energy tradeoff: AceSpec increases the tree size to mitigate network fluctuations and reduces the tree during favorable network conditions to conserve edge energy.

\subsubsection{Task-Aware Shape Allocation}
Given the network-aware budget $B^*(t)$, the system must determine the topological shape of $\mathcal{T}$. In our asymmetric tree architecture, the SLM generates a main draft chain of length $\gamma$. The remaining available budget for alternative branches is thus $B_{alloc}(t) = B^*(t) - \gamma$. 

Let $\mathbf{F} = [F_0, F_1, \dots, F_{\gamma-1}]$ denote the branch allocation vector, where $F_k \ge 0$ represents the number of alternative branches placed at depth $k$. The total allocation must satisfy the hardware constraint: $\sum_{k=0}^{\gamma-1} F_k \le B_{alloc}(t)$.

At any depth $k$, the probability that the cloud's target token is captured by the edge's cache relies on the output distribution of the language model. Let $\alpha$ denote the primary acceptance rate of the main chain token. If the main token is rejected, the target token is drawn from a residual distribution. Let $R(F_k)$ denote the cumulative probability that the target token falls within the top $F_k$ alternative candidates. The hit probability at depth $k$ is formulated as $p(F_k) = \alpha + (1-\alpha)R(F_k)$. 

Assuming conditional independence across sequence steps (a standard convention in speculative decoding analysis for mathematical tractability), the overall tree cache hit rate is the product of the hit probabilities at each depth:
\begin{equation}
    P_{hit}^{tree}(\mathbf{F}) = \prod_{k=0}^{\gamma-1} p(F_k) = \prod_{k=0}^{\gamma-1} \big( \alpha + (1-\alpha)R(F_k) \big)
    \label{eq:phit}
\end{equation}

By taking the logarithm of Eq. \eqref{eq:phit} to convert the product into a computationally tractable sum, we formulate the topological shape design as a constrained nonlinear optimization problem:
\begin{align}
    \max_{\mathbf{F}} \quad & \sum_{k=0}^{\gamma-1} \log \big( \alpha + (1-\alpha)R(F_k) \big) \label{eq:obj} \\
    \text{s.t.} \quad & \sum_{k=0}^{\gamma-1} F_k \le B_{alloc}(t) \label{eq:constraint1} \\
    & F_k \ge 0, \quad \forall k \in \{0, 1, \dots, \gamma-1\} \label{eq:constraint2}
\end{align}

Equation \eqref{eq:obj} represents the task-aware optimization goal, constrained by the network-aware limit in Eq. \eqref{eq:constraint1}. Solving this formulation yields the optimal non-uniform geometric decay strategy.

\subsection{Lagrangian Solution and Geometric Decay Allocation}

To solve the constrained optimization problem, we must account for the sequential dependency of the auto-regressive verification. In a speculative chain, a validation event at depth $k$ occurs only if all preceding $k-1$ tokens have been successfully accepted. Therefore, the prefix survival probability is $\alpha^k$. The effective expected log-likelihood objective is thus weighted by this survival decay:

\begin{equation}
    \mathcal{J}(\mathbf{F}) = \sum_{k=0}^{\gamma-1} \alpha^k \log \big( \alpha + (1-\alpha)R(F_k) \big)
    \label{eq:expected_log_likelihood}
\end{equation}

To derive a closed-form solution, we use the statistical properties of Large Language Models. Empirical analyses of LLM vocabulary distributions reveal a pronounced long-tail phenomenon. The residual sampling distribution exhibits a power-law sharpness, where the cumulative probability function can be bounded by $R(F_k) \approx 1 - c F_k^{-\beta}$, with $c > 0$ as a scaling constant and $\beta > 0$ representing the power-law decay exponent.

Substituting this into the hit probability yields $p(F_k) \approx 1 - c(1-\alpha)F_k^{-\beta}$. For high-confidence drafting, $p(F_k) \to 1$, allowing the approximation $\log p(F_k) \approx -c(1-\alpha)F_k^{-\beta}$. We then construct the Lagrangian function $\mathcal{L}(\mathbf{F}, \lambda)$ with the multiplier $\lambda \ge 0$ for the budget constraint:

\begin{equation}
    \mathcal{L}(\mathbf{F}, \lambda) = \sum_{k=0}^{\gamma-1} -c \alpha^k (1-\alpha) F_k^{-\beta} - \lambda \Big( \sum_{k=0}^{\gamma-1} F_k - B_{alloc}(t) \Big)
    \label{eq:lagrangian}
\end{equation}

By taking the partial derivative of $\mathcal{L}$ with respect to each $F_k$ and setting it to zero ($\frac{\partial \mathcal{L}}{\partial F_k} = 0$), we obtain the optimality condition:

\begin{equation}
\begin{split}
c \beta (1-\alpha) \alpha^k F_k^{-(\beta+1)} - \lambda &= 0 \\
\implies \quad F_k^* &= \left( \frac{c \beta (1-\alpha)}{\lambda} \right)^{\frac{1}{\beta+1}} \cdot \left( \alpha^{\frac{1}{\beta+1}} \right)^k
\end{split}
\label{eq:derivative}
\end{equation}

Let $\Phi(\lambda) = \big( \frac{c \beta (1-\alpha)}{\lambda} \big)^{\frac{1}{\beta+1}}$ be the depth-independent allocation base, and define the topological decay factor as $\eta = \alpha^{\frac{1}{\beta+1}}$. The closed-form solution for the optimal tree shape is given by:

\begin{equation}
    F_k^* = \Phi(\lambda) \cdot \eta^k
    \label{eq:fk_star}
\end{equation}

Since the model acceptance rate $\alpha \in (0, 1)$ and the power-law exponent $\beta > 0$, the decay factor satisfies $\eta \in (0, 1)$. To satisfy the budget constraint $\sum_{k=0}^{\gamma-1} F_k^* = B_{alloc}(t)$, we solve for $\Phi(\lambda)$ using the geometric series sum formula, yielding the final branch allocation at depth $k$:

\begin{equation}
    F_k^* = B_{alloc}(t) \cdot \frac{1 - \eta}{1 - \eta^\gamma} \cdot \eta^k
    \label{eq:fk_final}
\end{equation}

This formulation yields a non-uniform geometric decay strategy. It reveals a key insight: because sequential drafts have a decaying probability of verification, the system should not distribute the budget $B_{alloc}(t)$ uniformly. Instead, it allocates more branches at shallow depths ($k=0, 1$) where the pipeline is most vulnerable to early rejections. As depth $k$ increases, the probability of reaching that depth decreases exponentially, so the optimal resource allocation decays geometrically ($\eta^k$). This algorithm aligns the network budget constraints with model probability bounds, deploying computational redundancy where rejections are most likely to occur.

\section{Evaluation}
\label{sec:evaluation}

\begin{table*}[t]
\centering
\caption{End-to-End Throughput Speedup}
\label{tab:end_to_end}
\resizebox{\textwidth}{!}{%
\begin{tabular}{l ccc ccc ccc}
\toprule
\multirow{2}{*}{\textbf{Method}} & \multicolumn{3}{c}{\textbf{Qwen 0.6B \& 32B}} & \multicolumn{3}{c}{\textbf{Qwen 1.7B \& 32B}} & \multicolumn{3}{c}{\textbf{LLaMA 1B \& 70B}} \\
\cmidrule(lr){2-4} \cmidrule(lr){5-7} \cmidrule(lr){8-10}
& \textbf{GSM8K} & \textbf{HumanEval} & \textbf{Alpaca} & \textbf{GSM8K} & \textbf{HumanEval} & \textbf{Alpaca} & \textbf{GSM8K} & \textbf{HumanEval} & \textbf{Alpaca} \\
\midrule
Autoregressive & 1.00$\times$ (13.89) & 1.00$\times$ (13.91) & 1.00$\times$ (13.82) & 1.00$\times$ (13.89) & \textbf{1.00$\times$} (13.91) & 1.00$\times$ (13.82) & 1.00$\times$ (6.87) & 1.00$\times$ (6.86) & 1.00$\times$ (6.87) \\
Vanilla Spec.  & 0.57$\times$ & 0.44$\times$ & 0.41$\times$ & 0.51$\times$ & 0.49$\times$ & 0.42$\times$ & 1.35$\times$ & 1.45$\times$ & 0.81$\times$ \\
Split Inf.     & 0.54$\times$ & 0.53$\times$ & 0.51$\times$ & 0.54$\times$ & 0.53$\times$ & 0.51$\times$ & 0.51$\times$ & 0.66$\times$ & 0.65$\times$ \\
PicoSpec       & 1.45$\times$ & 1.13$\times$ & 1.02$\times$ & 1.17$\times$ & \textbf{1.00$\times$} & 0.94$\times$ & 2.51$\times$ & 2.90$\times$ & 2.26$\times$ \\
DSSD           & 0.85$\times$ & 0.74$\times$ & 0.34$\times$ & 0.44$\times$ & 0.32$\times$ & 0.27$\times$ & 1.80$\times$ & 1.95$\times$ & 1.27$\times$ \\
\textbf{AceSpec (Ours)} & \textbf{1.75$\times$} & \textbf{1.18$\times$} & \textbf{1.32$\times$} & \textbf{1.38$\times$} & 0.98$\times$ & \textbf{1.16$\times$} & \textbf{3.29$\times$} & \textbf{3.52$\times$} & \textbf{3.13$\times$} \\
\bottomrule
\end{tabular}%
}
\end{table*}

\subsection{Experimental Setup}
\label{subsec:setup}

We implement AceSpec and all baseline frameworks using PyTorch and evaluate them on a simulated edge-cloud testbed. The edge client runs on an NVIDIA Jetson AGX Orin (32GB), representing a typical resource-constrained but parallel-capable edge node. The cloud server is equipped with 4$\times$ NVIDIA A100 (40GB) GPUs to host the target LLMs. To replicate real-world Wide Area Network (WAN) conditions, we use the Linux Traffic Control (\texttt{tc}) utility to constrain the network throughput and inject latency, with the baseline bandwidth restricted to 100 Mbps. 

To evaluate AceSpec across different model architectures and parameter scales, we select model pairs from the Qwen and LLaMA families. Specifically, we construct the Qwen-0.6B/32B and Qwen-1.7B/32B collaborative pairs, alongside the LLaMA-1B/70B pair. The end-to-end evaluations use three datasets with varying generative complexities and token acceptance rates: GSM8K for multi-step mathematical reasoning, HumanEval for code generation, and Alpaca for general instruction-following tasks.

We compare AceSpec against five inference paradigms. The first is standard Autoregressive sequential decoding, which acts as a cloud-only baseline without edge speculation. The second is Vanilla Speculative Decoding (Vanilla Spec.), which deploys standard speculative decoding over the WAN using a synchronous ``stop-and-wait'' mechanism. We also evaluate Split Inference (Split Inf.), a collaborative paradigm where early model layers execute on the edge device and the remaining layers are processed on the cloud. Furthermore, we compare against two latest cloud-edge collaborative speculated decoding frameworks: DSSD\cite{ning2025dssd} (Distributed Split Speculative Decoding), which offloads partial verification to the edge to reduce communication overhead, and PicoSpec\cite{zhang2026pipelined}, an asynchronous pipeline framework relying on single-chain drafting. 

\subsection{End-to-End Performance Comparison}
\label{subsec:end_to_end}

Table \ref{tab:end_to_end} presents the throughput speedup of AceSpec and the baseline methods across various model pairs and datasets, normalized against the Autoregressive baseline. Under constrained WAN conditions, traditional collaborative paradigms such as Vanilla Speculative Decoding and Split Inference experience substantial performance degradation, frequently yielding speedup ratios below 1.00x. This phenomenon highlights the impact of the communication wall, where the transmission latency of uncompressed tensors or hidden states offsets the theoretical benefits of edge-cloud parallelization. In contrast, AceSpec achieves state-of-the-art performance across most configurations. For instance, with the LLaMA-1B and 70B model pair on the HumanEval dataset, AceSpec achieves a 3.52x throughput speedup. Similarly, using the Qwen-0.6B and 32B configuration on the GSM8K dataset, AceSpec delivers a 1.75x speedup, outperforming the single-chain asynchronous pipeline of PicoSpec (1.45x) and the distributed architecture of DSSD (0.85x). This performance margin demonstrates the efficacy of the asymmetric multi-branch caching mechanism in neutralizing the rollback penalty associated with nucleus sampling rejections.

Despite its overall performance, we observe a specific degradation with the Qwen-1.7B and 32B model pair on the HumanEval dataset, where AceSpec yields a speedup of 0.98x, slightly underperforming the cloud-only Autoregressive baseline (1.00x) and PicoSpec (1.00x). This regression stems from the interaction between the draft model's parameter scale and the characteristics of code generation tasks. HumanEval demands rigid syntactic precision, often leading to highly peaked target distributions and frequent rejections if the draft deviates structurally. When employing a larger draft SLM (1.7B parameters) on the resource-constrained edge device, the local computational latency required to expand a multi-branch token tree increases. Consequently, the edge generation time begins to exceed the cloud's turnaround time ($T_{tree\_draft} > T_{cloud}$). The asynchronous pipeline becomes compute-bound at the edge rather than network-bound, causing the overhead of local multi-branch drafting to outweigh the latency-hiding benefits. In contrast, when using the lighter Qwen-0.6B model on the identical task, the system maintains a positive acceleration (1.18x) because the smaller SLM is fast enough to complete its multi-branch expansion within the asynchronous window. This comparative degradation supports the theoretical boundaries established in our Network-Aware Budgeting constraint, emphasizing that the multi-branch cache expansion must be bounded to remain hidden within the network round-trip time.

\subsection{Robustness under Constrained Bandwidth}
\label{subsec:bandwidth}

\begin{figure}[h]
    \centering
    \includegraphics[width=\linewidth]{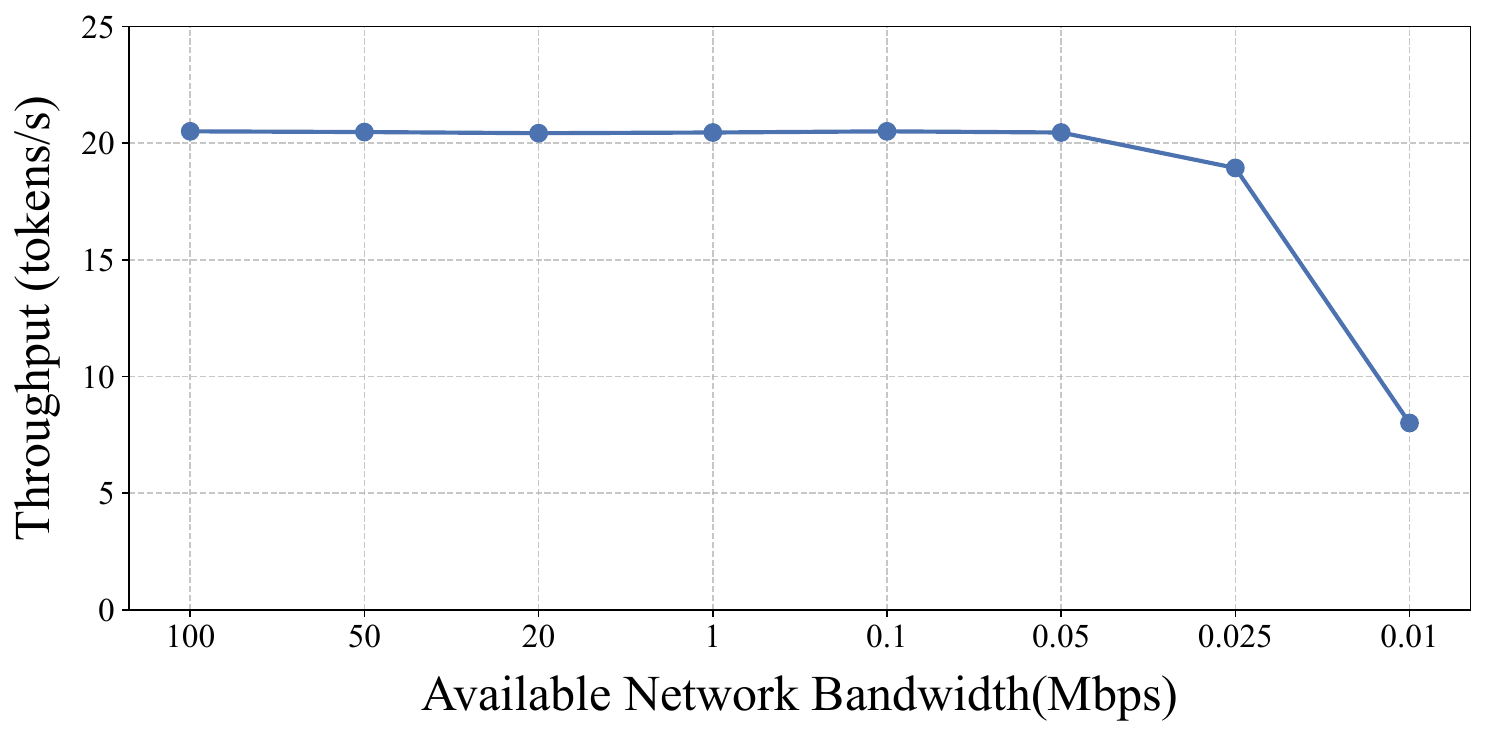}
    \caption{End-to-end performance of AceSpec under varying network bandwidths.}
    \label{fig:bandwidth}
\end{figure}

To evaluate the resilience of AceSpec against network volatility, we simulate a constrained and variable WAN environment. Fig. \ref{fig:bandwidth} illustrates the end-to-end performance of AceSpec as the available network bandwidth decreases from 100 Mbps to 10 Kbps.

AceSpec exhibits ``bandwidth immunity'' across various network conditions. As the bandwidth drops from 100 Mbps to 50 Kbps, the throughput remains unaffected, maintaining a stable plateau. This robustness validates the efficacy of the asymmetric transmission protocol. By transmitting only a compact array of main-chain token indices over the uplink and returning a sparsified target distribution via the downlink, AceSpec minimizes the communication payload. Consequently, the pipeline avoids the communication bottlenecks that typically degrade traditional collaborative frameworks under sub-megabit conditions.

Performance degradation occurs only when the bandwidth is restricted to 25 Kbps and 10 Kbps. At these thresholds, the transmission delay of the minimal scalar data begins to exceed the computational latency of the edge device. Despite this drop, maintaining near-peak performance at 50 Kbps demonstrates the practical viability of AceSpec for deployment in unstable or remote edge environments.

\subsection{Ablation Study on Branch Allocation Strategies}
\label{subsec:ablation}

To evaluate the efficacy of the Lagrangian-optimized non-uniform branch allocation strategy, we conduct an ablation study comparing three drafting topologies. As presented in Table \ref{tab:ablation}, we evaluate a Single Branch configuration (analogous to traditional pipelined frameworks), a Uniform Multi-Branch configuration where the computational budget is distributed evenly across all depth levels, and the proposed Non-Uniform Branch configuration.

\begin{table}[h]
\centering
\caption{Ablation Study on Different Branch Allocation Strategies. Metrics include TTFT (ms), Draft Hit Rate (\%), and Throughput (tokens/s).}
\label{tab:ablation}
\resizebox{\linewidth}{!}{%
\begin{tabular}{l l c c c}
\toprule
\textbf{Method} & \textbf{Dataset} & \textbf{TTFT (ms)} $\downarrow$ & \textbf{Hit Rate} $\uparrow$ & \textbf{Throughput} $\uparrow$ \\
\midrule
\multirow{3}{*}{Single Branch} 
 & GSM8K & 442.48 & 0.50 & 20.14 \\
 & HumanEval & 173.30 & 0.77 & 15.17 \\
 & Alpaca & 175.14 & 0.23 & 14.09 \\
\midrule
\multirow{3}{*}{Uniform Branch} 
 & GSM8K & 450.79 & 0.94 & 21.58 \\
 & HumanEval & 188.36 & 0.91 & 15.61 \\
 & Alpaca & 174.01 & 0.74 & 14.24 \\
\midrule
\multirow{3}{*}{\textbf{Non-Uniform Branch}} 
 & GSM8K & \textbf{436.62} & \textbf{0.97} & \textbf{24.37} \\
 & HumanEval & \textbf{171.33} & \textbf{0.97} & \textbf{16.57} \\
 & Alpaca & \textbf{173.21} & \textbf{0.87} & \textbf{16.59} \\
\bottomrule
\end{tabular}%
}
\end{table}

The results show that the non-uniform strategy outperforms the others across all evaluated datasets. In terms of overall generation efficiency, the non-uniform allocation achieves the highest throughput, reaching 24.37 tokens/s on the GSM8K dataset, compared to 20.14 tokens/s for the single branch and 21.58 tokens/s for the uniform topology. This gain confirms our theoretical derivation: uniformly expanding deep nodes yields diminishing marginal returns due to compounded rejection probabilities. Shifting the branch budget toward shallower, high-confidence nodes maximizes the cache hit rate under a fixed computational budget.

Analyzing the latency and hit rate components provides insights into edge compute dynamics. The Uniform Multi-Branch strategy frequently increases the Time to First Token (TTFT) (e.g., from 442.48 ms to 450.79 ms on GSM8K) because constructing a dense, uniform tree saturates the compute capacity of the edge and delays the initial dispatch of the main chain. In contrast, the Non-Uniform strategy prunes the expansion space, mitigating the initial latency overhead (achieving 436.62 ms TTFT on GSM8K) while significantly improving the draft hit rate to 0.97. We note a minor anomaly in the Alpaca dataset, where the Uniform strategy exhibits a marginally lower TTFT compared to the Single Branch baseline. Given the minimal difference of less than 2 ms, this fluctuation is primarily attributed to standard system variance and CUDA kernel launch overheads on the edge device under lightweight prompts. By concentrating the parallel capacity of the draft SLM on statistically viable alternative paths, AceSpec translates its mathematical hit-rate optimization into end-to-end acceleration.

Finally, it is worth acknowledging that the multi-branch drafting mechanism incurs certain computational overhead in its current implementation. Future optimizations, such as deploying more efficient fused attention kernels and advanced memory management, could further minimize this overhead and elevate the end-to-end inference performance.

\subsection{Sensitivity Analysis}
\label{subsec:sensitivity}

To understand the operational boundaries and robustness of AceSpec, we conduct a sensitivity analysis on two hyperparameters: the edge computational budget ($B$) and the nucleus sampling temperature ($T$). The draft hit rate serves as the primary metric for this analysis because it determines the efficiency of the $\mathcal{O}(1)$ state recovery.

\begin{figure}[h]
    \centering
    \includegraphics[width=\linewidth]{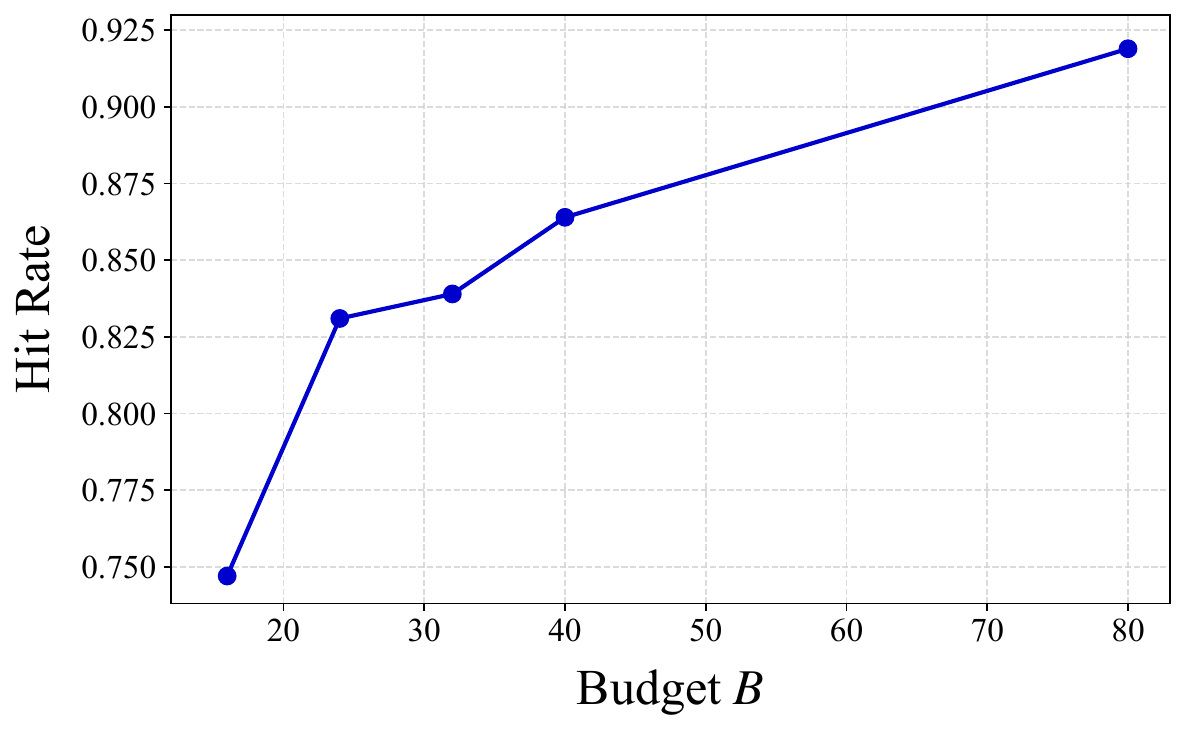}
    \caption{Draft hit rate as a function of the branch budget $B$. The curve demonstrates diminishing marginal returns, guiding the allocation of edge compute.}
    \label{fig:budget}
\end{figure}

Fig. \ref{fig:budget} illustrates the draft hit rate as a function of the branch budget $B$, which represents the maximum number of alternative tokens the edge SLM explores during the asynchronous window. The results exhibit a non-linear trend: as the budget increases from 20 to 40, the hit rate rises rapidly from approximately 0.750 to over 0.850. As $B$ scales toward 80, the growth flattens, indicating diminishing marginal returns. This phenomenon aligns with the heavy-tailed distribution of natural language; allocating a small budget captures the highly probable tokens, whereas expanding the tree further only covers the low-probability tail. This insight provides a basis for selecting a budget threshold that maximizes the hit rate while preventing compute saturation at the edge.

\begin{figure}[h]
    \centering
    \includegraphics[width=\linewidth]{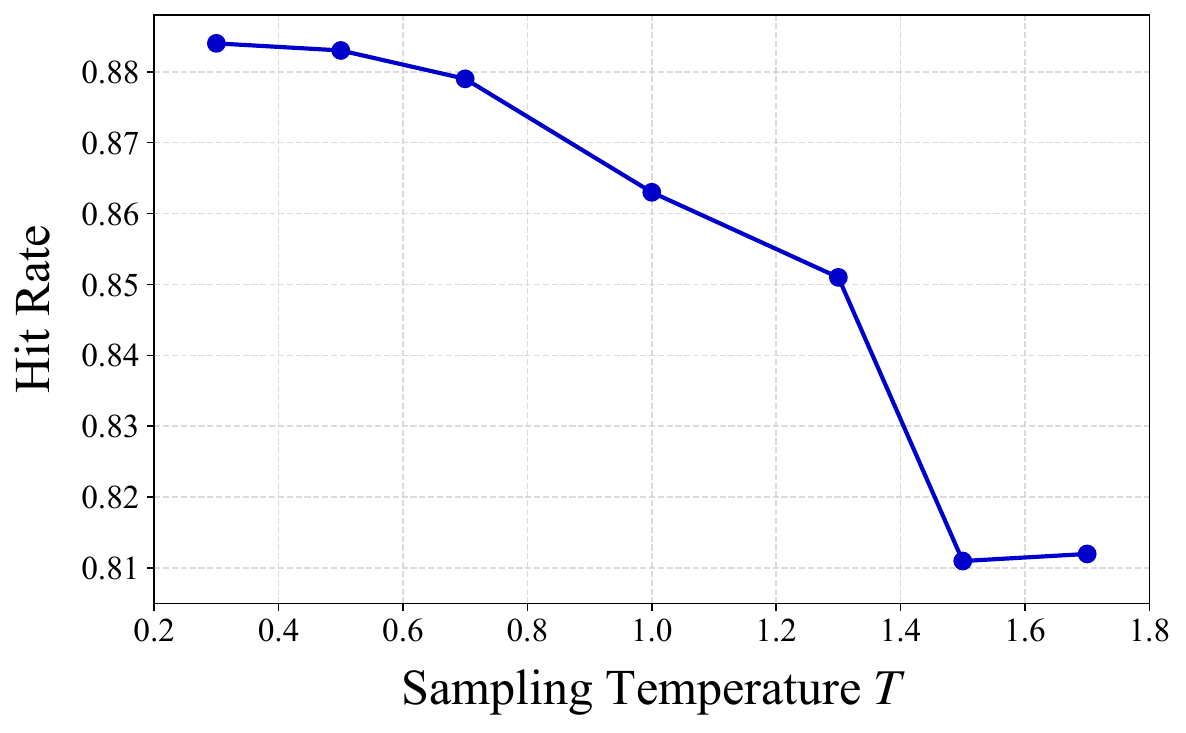}
    \caption{Comparison of draft hit rates under different sampling temperatures $T$. AceSpec maintains a hit rate above 81\% even in diverse generation scenarios.}
    \label{fig:tem}
\end{figure}

In Fig. \ref{fig:tem}, we analyze the resilience of the system to generation randomness by varying the sampling temperature $T$ from 0.3 to 1.7. Elevating the temperature flattens the target distribution, increasing decoding diversity and lowering the baseline acceptance rate. Consequently, the cache hit rate declines from 0.885 at low temperatures to 0.811 at high temperatures ($T \ge 1.5$). The drop becomes steeper when the temperature exceeds 1.0. Even with this increased randomness, AceSpec maintains a hit rate above 81\%. This indicates that the multi-branch cache captures a broad set of alternative tokens, demonstrating that AceSpec is effective for both deterministic reasoning and creative text generation tasks.

\section{Conclusion}
\label{sec:conclusion}

In this paper, we propose AceSpec, an asymmetric edge-cloud collaborative framework designed to overcome communication bottlenecks and pipeline stalls in distributed Large Language Model inference. By combining asynchronous multi-branch expansion, an $\mathcal{O}(1)$ probabilistic KV cache, and a Lagrangian-optimized non-uniform branch allocation strategy, AceSpec masks WAN latency and eliminates redrafting penalties. Extensive evaluations demonstrate that AceSpec achieves up to a 3.52$\times$ throughput speedup over traditional baselines while maintaining bandwidth immunity under constrained network conditions (down to 50 Kbps). This provides an efficient, scalable, and robust solution for deploying LLMs in volatile edge environments.

\bibliographystyle{IEEEtran}

\bibliography{ref.bib}

\end{document}